\documentclass[aps,prl,twocolumn,superscriptaddress,nofootinbib,floatfix]{revtex4-2}

\usepackage{amsmath}
\usepackage{amssymb}
\usepackage{graphicx}
\usepackage{bm}
\usepackage{xcolor}
\usepackage{listings}
\usepackage[colorlinks=true,allcolors=blue]{hyperref}

\newcommand{\Z}{\mathbb{Z}}
\newcommand{\RP}{\mathbb{RP}}
\newcommand{\Spin}{\mathrm{Spin}}
\newcommand{\Pin}{\mathrm{Pin}}
\newcommand{\Om}{\Omega}
\newcommand{\Neq}[1]{\mathcal{N}=#1}

\newcommand{\code}[1]{\texttt{#1}}
\graphicspath{{figs/}}

\definecolor{kw}{RGB}{0,0,180}
\definecolor{cmt}{RGB}{110,110,110}
\definecolor{str}{RGB}{160,60,20}
\begin{document}

\title{Anomaly Matching between Topological Superconductors and $\mathcal{N}=8$ Supergravity}

\author{Christopher~W.~Murphy}
\email{murphtron5000@proton.me}
\affiliation{Klaviyo, Boston, MA 02111, USA}

\date{\today}

\begin{abstract}
The boundary of a $3{+}1$-dimensional topological superconductor carries a $\Z_{16}$-valued global (Dai-Freed) anomaly, famously matched by the $16$ fermions of a Standard Model generation once a certain $\Z_4$ symmetry is gauged. 
In a supergravity completion, though, a gravitino contributes $-7/16$ rather than $\pm1/16$, obstructing this $\Z_4$ in the Minimal Supersymmetric Standard Model. 
We observe that four-dimensional $\Neq{8}$ supergravity evades this fate. 
Its fermions, eight gravitini in the $\mathbf 8$ and fifty-six spin-$\tfrac12$ states in the $\mathbf{56}=\Lambda^3\mathbf 8$ of the $SU(8)$ R-symmetry, cancel the anomaly. 
The cancellation holds for \emph{every} admissible $\Z_4$ structure, is consistent with the vanishing of the continuous $SU(8)$ anomaly, and follows from a Smith homomorphism and the Rarita-Schwinger ghost-subtraction weight identifying the supergravity anomaly with that of the topological superconductor.
In maximal supergravity this $\Z_4$ stays free of this particular obstruction to gauging despite its eight gravitini.
All numbers are reproduced by an open-source library described in the Supplemental Material.
\end{abstract}

\maketitle

%======================================================================
Anomalies are among our sharpest consistency conditions; any symmetry we wish to gauge must
be free of them. 
Beyond the familiar perturbative and global anomalies lies a further, more stringent requirement, cleanly organized by the Dai-Freed theorem~\cite{Dai:1994kq, Witten:2015aba, Yonekura:2016wuc}. 
For a $d$-dimensional theory of fermions on some manifold $X$ charged under symmetry group $G$, the partition function's phase is controlled by the exponentiated Atiyah-Patodi-Singer (APS) $\eta$-invariant of a Dirac operator one dimension up. 
When perturbative anomalies cancel, $\exp(2\pi i\,\eta_Y)$ is a bordism invariant, and anomaly freedom becomes the statement that
\begin{equation}
  \exp\!\left(2\pi i\,\eta_Y\right)=1
\end{equation} 
for all closed 
\begin{equation}
  Y\in \Om^{\text{structure}}_{d+1}(BG),
\end{equation}
where the manifolds $Y$ satisfy $X = \partial Y$ and share the same (s)pin structure as $X$, and $BG$ is the classifying space of $G$.
That is, the anomaly, viewed as a homomorphism $\Om^{\text{structure}}_{d+1}(BG)\to U(1)$, is trivial~\cite{GarciaEtxebarria:2018ajm}. 
These ``Dai-Freed anomalies'' can impose constraints invisible to the triangle diagram.

The paradigmatic example lives on the boundary of a $3{+}1$-dimensional topological superconductor~\cite{Witten:2015aba, Hsieh:2015xaa}. 
A single Majorana fermion there has a time-reversal anomaly valued in
\begin{equation}
  \Om^{\Pin^+}_4=\Z_{16},
\end{equation}
so consistency requires the number of Majorana modes to be a multiple of $16$. 
Strikingly, each generation of the Standard Model (SM) with a right-handed neutrino has exactly $16$ Weyl fermions.
In the Minimal Supersymmetric Standard Model (MSSM) the matter and gauge sector contributions to the anomaly cancel beautifully. 
However, the single gravitino in the MSSM spoils the cancelation.
As such, it is natural to ask whether \emph{any} theory of gravity gets the $16$ right.
Hypothetically, if there was a theory with 16 gravitini, it would trivially satisfy the anomaly condition.
Unfortunately, for $d = 4$ the maximal amount of supersymmetry is $\Neq{8}$, i.e.\ $8$ gravitini.
The purpose of this Letter is to show that $\Neq{8}$ supergravity in four dimensions does satisfy the anomaly condition in a non-trivial manner.

Let us start by making precise the connection between the topological superconductor and the SM by following the analysis of Garc\'ia-Etxebarria and Montero~\cite{GarciaEtxebarria:2018ajm}.
A Smith homomorphism~\cite{bahri1987, Kapustin:2014dxa}
\begin{equation}
  \Om^{\Spin^{\Z_4}}_5 \;\cong\; \Om^{\Pin^+}_4=\Z_{16} 
  \label{eq:smith}
\end{equation}
turns each four-dimensional Weyl fermion of $\Z_4$ charge $1$ into one three-dimensional $\Pin^+$ Majorana mode.
Ref.~\cite{Tachikawa:2018njr} defined $\Spin^{\Z_4} = (\Spin \times \Z_4)~/~\Z_2$ where $\Z_2$ is identified with fermion parity $(-1)^F$.
Candidates for the $\Z_4$ include operator dimension parity, the center of the four-dimensional conformal group, $Z(\Spin(4,2))$, and the center of the $SO(10)$ gauge group~\cite{Murphy:2024zfw}.
Concretely, $\RP^5$ generates $\Om_5^{\Spin^{\Z_4}}(\mathrm{point})=\Z_{16}$, so it is the only class one must check, and a Weyl fermion of $\Z_4$ charge $q$ on it contributes $\eta=\chi(q)/16$, where
\begin{equation}
  \chi(q)=\begin{cases}
    +1, & q\equiv 1,\\
    -1, & q\equiv 3,\\
    \phantom{+}0, & q\equiv 0,2,
  \end{cases}\pmod 4,
  \label{eq:chi}
\end{equation}
so a charge-$2$ fermion is inert and a charge-$3(=-1)$ fermion flips the sign. 
The anomaly of a spectrum is the integer
\begin{equation}
  \nu \;=\; \sum_i m_i\, w(\text{spin}_i)\,\chi(q_i)\ \ (\mathrm{mod}\ 16),
  \label{eq:nu}
\end{equation}
with multiplicities $m_i$; it must vanish for consistency.
The combination of hypercharge and baryon and lepton numbers, $X=-2Y+5(B{-}L)$, assigns every left-handed SM fermion $q \equiv X \bmod 4 = 3$ (and right-handed SM fermions have $X \bmod 4 = 1$), so its $\Z_4$ reduction furnishes a $\Spin^{\Z_4}$ structure.

The weight $w$ in Eq.~\eqref{eq:nu} is $1$ for a Weyl fermion but \emph{not} for a gravitino. 
A Rarita-Schwinger field is a vector-spinor minus a spin-$\tfrac12$ ghost of the opposite chirality. 
On $\RP^5$ the vector-spinor Dirac index is $6$ times the ordinary one, and subtracting the ghost adds one more, so~\cite{GarciaEtxebarria:2018ajm}
\begin{equation}
  w_{3/2}=6+1=7 .
  \label{eq:seven}
\end{equation} 
Equivalently, in the domain-wall picture of Ref.~\cite{GarciaEtxebarria:2018ajm} a gravitino localizes seven $\Pin^+$ Majorana modes on the wall, against one for a Weyl fermion. 
In the MSSM, the $12$ gauginos and four higgsinos all have $X \bmod 4$ even, so they do not contribute to the anomaly, and we retain the SM result.
Alternatively, operator dimension parity assigns the same charge to every canonically-dimensioned spin-$\tfrac12$ field regardless of its gauge quantum numbers.
In particular, under this charge the left(right)-handed $16$ SM fermions, $12$ gauginos and $4$ higgsinos all carry $\Z_4$ charge $3(1)$, and $16+12+4=32\equiv16k\equiv0$.
However, in both cases, a single gravitino adds $-7$, and $16k-7 \equiv 9\neq 0$.
Once gravity is dynamical the match is broken, motivating the search for a gravitational theory the satisfies the anomaly. 

%======================================================================

\begin{figure*}[t]
  \centering
  \includegraphics[width=\linewidth]{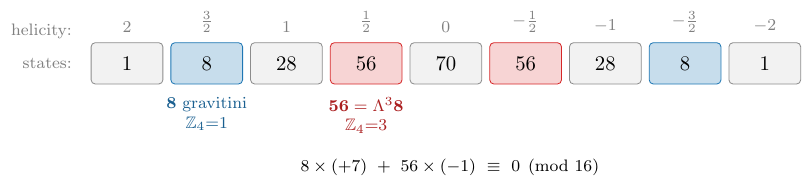}
  \caption{
    The $\Neq{8}$ helicity multiplet, $\binom{8}{k}$ states at helicity $2-k/2$. 
    The fermionic slots are the $\mathbf 8$ gravitini ($\Z_4$ charge $1$) and the $\mathbf{56}=\Lambda^3\mathbf 8$ spin-$\tfrac12$ fermions ($\Z_4$ charge $3$).
  }
  \label{fig:diamond}
\end{figure*}

Now consider $\Neq{8}$ supergravity in four dimensions~\cite{Cremmer:1979up, Cremmer:1978km, FreedmanVanProeyen}. 
Its massless multiplet contains $\binom{8}{k}$ states of helicity $2-k/2$. 
The fermions sit at $k=1$ and $k=3$, which is illustrated in Fig.~\ref{fig:diamond}.
They transform in the $\mathbf 8$ and the three-fold antisymmetric $\mathbf{56}$ of the $SU(8)$ R-symmetry, 
\begin{equation}
  \underbrace{\mathbf 8}_{8\ \text{gravitini}}\ (h=\tfrac32),
  \quad
  \underbrace{\mathbf{56}=\Lambda^3\mathbf 8}_{56\ \text{spin-}\frac12}\ (h=\tfrac12),
\end{equation}
on which $SU(8)$ acts \emph{chirally}~\cite{FreedmanVanProeyen}. 
This chiral action is exactly what a $\Spin^{\Z_4}$ structure needs. 
The center of $SU(8)$ contains a $\Z_4$ generated by $g=\mathrm{diag}(i,\dots,i)$, with $g^2=-\mathbf 1=(-1)^F$ on the fermions; it assigns the $\mathbf 8$ charge $1$ and hence the $\mathbf{56}=\Lambda^3\mathbf 8$ charge $3\cdot1\equiv 3\pmod 4$. 
Since $\det g=i^8=1$, this $\Z_4$ lies in $SU(8)$. 
It also sits in the intersection with the overall $U(1)$ of the R-symmetry, i.e.\ in the $\Z_8$ center of $SU(8)$.
Although this $SU(8)$ is a composite (local) $R$-symmetry, the $\Z_4$ acts faithfully on the fermions with $g^2=(-1)^F$, so it defines a genuine $\Spin^{\Z_4}$ background under which to place the theory.
The invariant $\nu$ is then the 't~Hooft obstruction to formulating the theory consistently on backgrounds with this topology.
This is true whether the $\Z_4$ is ultimately realized as an independently gauged symmetry or only as part of the composite $SU(8)$ connection.
A first-principles derivation of which of these applies is left for future work.
However, Debray and Yu~\cite{Debray:2022wcd} have already carried out the analogous derivation for the full continuous $SU(8)$ (equivalently the U-duality group $E_{7(7)}$ of $\Neq{8}$ supergravity), which is a genuine, if composite, symmetry of the theory.\footnote{
Placing 4d $\Neq{8}$ supergravity on $\Spin\text{-}SU(8) = (\Spin \times SU(8))~/~\Z_2$ manifolds with the same $\Z_2=(-1)^F$ identification we use, but applied to the whole of $SU(8)$ rather than to our $\Z_4$.
They compute $\Omega_5^{\Spin\text{-}SU(8)}\cong\Z/2$ via the Adams spectral sequence and show, by an independent $\eta$-invariant computation on the Wu manifold, that this $\Z/2$-valued anomaly of the actual $\mathbf 8\oplus\mathbf{28}\oplus\mathbf{56}$ matter content vanishes.
}
Since any $\Z_4\subset SU(8)$ with $g^2=(-1)^F$ (including every embedding in Table~\ref{tab:enum}) reduces this structure, their result should already force our $\nu$ to vanish via the induced map of bordism categories.
However, a computation of the induced map would be needed to confirm this.
We regard our $\Z_{16}$-valued computation as an independent, logically distinct, and finer-grained verification, tied specifically to the topological superconductor/Standard Model correspondence of Eq.~\eqref{eq:smith}, rather than a first demonstration that this matter content is anomaly-free.
Equivalently, on this spectrum the $SU(8)$-center $\Z_4$ coincides with operator dimension parity since the charge $4j_l-2d$ $\bmod 4$ assigns the gravitini ($(1,\tfrac12)$, $d=\tfrac32$) charge $1$ and the spin-$\tfrac12$ fermions ($(\tfrac12,0)$, $d=\tfrac32$) charge $3$. 
Equation~\eqref{eq:main} extends operator dimension parity, whose mod-16 anomaly is known for spin-$\tfrac12$ matter~\cite{Murphy:2024zfw, GarciaEtxebarria:2018ajm}, to the gravitino via the Rarita-Schwinger value of Eq.~\eqref{eq:seven}.
Only fermions enter Eq.~\eqref{eq:nu} (the graviton, $28$ vectors and $70$ scalars are bosonic), so
\begin{equation}
  \boxed{\ \nu_{\Neq{8}}
  = \underbrace{8\times(+7)}_{\text{gravitini}}
  + \underbrace{56\times(-1)}_{\text{spin-}\frac12}
  = 0.\ }
  \label{eq:main}
\end{equation}
$\Neq{8}$ supergravity is free of the topological-superconductor anomaly. 
For the content $\{\mathcal{N}\text{ gravitini},\ \binom{\mathcal{N}}{3}\text{ spin-}\tfrac12\}$, which is the full multiplet at $\mathcal{N}\le4$ and $\mathcal{N}=8$, $\nu=0$ requires $7\mathcal{N}=\binom{\mathcal{N}}{3}$ at the central embedding, whose only solutions are $\mathcal{N}=\{0,8\}$. The eight gravitini punch seven times above their weight and exactly cancel the fifty-six oppositely charged spin-$\tfrac12$ fermions, no generic consequence of supersymmetry. 
Via Eq.~\eqref{eq:smith} this is precisely an anomaly matching across the Smith homomorphism, a correspondence between the $\Z_{16}$ boundary anomaly of the topological superconductor and the (vanishing) anomaly class of maximal supergravity.

%======================================================================
One might worry that Eq.~\eqref{eq:main} depends on the particular $\Z_4$ chosen. 
It does not. 
Any $\Z_4\subset SU(8)$ with $\Z_2=(-1)^F$ is generated by an element with $g^2=-\mathbf 1$, i.e.\ $p$ eigenvalues $+i$ and $8-p$ eigenvalues $-i$ on the $\mathbf 8$; $\det g=1$ forces $p$ even. 
The $\mathbf{56}$ then splits into components of charge $9-2a\pmod4$ with multiplicity $\binom{p}{a}\binom{8-p}{3-a}$. 
As shown in Table~\ref{tab:enum}, summing Eq.~\eqref{eq:nu} over all $\binom{8}{3}$ components gives $\nu=0$ for every case. 
Furthermore, at $p=0,8$ the cancellation is exact as unreduced integers ($56-56=0$), a stronger and more delicate statement than the mod-$16$ vanishing that consistency actually requires. 
Maximal supergravity is anomaly-free on all admissible $\Spin^{\Z_4}$ backgrounds, not merely the central one. 
(A fixed $\Z_4$ gives a single condition on the generator $\RP^5$; Table~\ref{tab:enum} instead varies the embedding of the $\Z_4$.)

\begin{table}[b]
  \caption{The mod-$16$ anomaly $\nu$ for every $\Z_4\subset SU(8)$ with $\Z_2=(-1)^F$,
  labeled by $p$ (the number of $+i$ eigenvalues on the $\mathbf 8$).}
  \label{tab:enum}
  \begin{ruledtabular}
  \begin{tabular}{ccccc}
   $p$ & gravitini & spin-$\tfrac12$ & $\nu \pmod{16}$ & \\
   \hline
   $0$ & $-56$ & $+56$ & $0$ & \checkmark\\
   $2$ & $-28$ & $-4$  & $0$ & \checkmark\\
   $4$ & $0$   & $0$   & $0$ & \checkmark\\
   $6$ & $+28$ & $+4$  & $0$ & \checkmark\\
   $8$ & $+56$ & $-56$ & $0$ & \checkmark\\
  \end{tabular}
  \end{ruledtabular}
\end{table}

Ungauged $\Neq{8}$ supergravity has an abelian gauge group (twenty-eight $U(1)$'s).
With no instantons there is no mixed non-abelian $\Z_4$--gauge anomaly of the type that afflicts the MSSM, where an $S^1\times S^4$ computation exposes $SU(2)$-- and $SU(3)$--$\Z_4$ anomalies~\cite{GarciaEtxebarria:2018ajm}.
Only $\mathbf 8\oplus\mathbf{56}$ enter Eq.~\eqref{eq:main}: the $\mathbf{28}=\Lambda^2\mathbf8$ carries $\Z_4$ charge $2$ (since $g$ acts on $\Lambda^k\mathbf8$ as $i^k$), so it is inert under our $\Z_4$ ($\chi(2)=0$) regardless of its bosonic statistics.
The relevant continuous anomaly is instead that of the composite $SU(8)$ itself, computed long ago by Marcus~\cite{Marcus:1985yy}, with cubic-Casimir ratios $1:4:5$ for the $\mathbf 8,\mathbf{28},\mathbf{56}$ and helicity weights $-3,+2,-1$,
\begin{equation}
  (-3)\cdot 1 + (2)\cdot 4 + (-1)\cdot 5 = 0,
\end{equation}
which, unlike Eq.~\eqref{eq:main}, does receive a contribution from the $\mathbf{28}$.
Vanishing of this continuous $SU(8)$ anomaly does not by itself imply that a discrete $\Z_4\subset SU(8)$ is anomaly-free: a discrete subgroup of an anomaly-free continuous symmetry can still carry an independent Dai-Freed anomaly, which is the entire premise of the program used here.
We therefore record the discrete result Eq.~\eqref{eq:main} and Marcus' continuous one~\cite{Marcus:1985yy} as mutually consistent, both ultimately rooted in the same $\mathbf 8\oplus\mathbf{56}$ chiral fermion content (with the $\mathbf{28}$ additionally entering the continuous computation), rather than one implying the other.
Marcus' result is perturbative, i.e. local.
The corresponding \emph{non}-perturbative statement for the full continuous $SU(8)$, including the $\mathbf{28}$, is precisely what Debray and Yu~\cite{Debray:2022wcd} establish.
A full treatment of \emph{gauged} $\Neq{8}$~\cite{deWit:1982bul} would additionally require the gravitino Rarita-Schwinger index in a non-abelian instanton background and is left open.

The vanishing is special, not an automatic consequence of $\Neq{8}$ being a consistent theory. 
Applying the same $\Z_4=\mathrm{diag}(i,\dots,i)$ to $\mathcal{N} < 8$ pure supergravity ($\mathcal{N}$ gravitini in the $\mathbf{N}$ and $\binom{\mathcal{N}}{3}$ spin-$\tfrac12$ fermions in $\Lambda^3\mathbf{N}$) gives $\nu=7,14,4,8$ for $\mathcal{N}=1,2,3,4$, all nonzero, and it also fails once the MSSM gravitino is included.\footnote{
Note that $\det g=i^{\mathcal N}=1$ only for $\mathcal N\equiv0\pmod4$, so this $g$ sits in $SU(\mathcal N)$ only at $\mathcal N=4$.
For $\mathcal N=1,2,3$ it instead lies along the anomalous overall $U(1)_R$, and those three data points illustrate the arithmetic pattern of Eq.~\eqref{eq:nu} rather than genuine $SU(\mathcal N)$-embedded $\Z_4$ tests, for which one would need an $\mathcal N$-appropriate generator with $g^2=(-1)^F$ and $\det g=1$.
}
These are perfectly consistent theories, so $\nu$ obstructs not their existence but the gauging of \emph{this particular} $\Z_4$.
Among the cases we check it cancels only for maximal $\Neq{8}$ (and the SM), where the $\Z_4$ can be gauged. This is consistent with Marcus' continuous result, as far as we have checked ($\mathcal{N}\le4$ and $\mathcal{N}=8$; $\mathcal{N}=7$ is equivalent to $\Neq{8}$, while $\mathcal{N}=5,6$ involve additional fermions from CPT completion whose $R$-charge assignment we do not detail here).

The weight $w_{3/2} = 7$ of Eq.~\eqref{eq:seven} is fixed by the Rarita-Schwinger index theorem~\cite{GarciaEtxebarria:2018ajm, Alvarez-Gaume:1983ihn}.
The weight applies to each of the eight $\Neq{8}$ gravitini, all carry $\Z_4$ charge $1$, so each contributes $+7$, the $SU(8)$ representation entering only through this charge.
Moreover $\nu$ is a bordism invariant of the free operators in the $\Spin^{\Z_4}$ background, so it is blind to the multiplet's graviphoton and Yukawa couplings; on $\RP^5$ no graviphoton or $SU(8)$ bundle is switched on, so mixed contributions (which would live in $\Om_5^{\Spin^{\Z_4}}(BG_R)$, part of the gauged analysis left open above) do not arise. 

%======================================================================
In maximal supergravity this $\Z_4$ is anomaly-free and can be gauged, whereas in the MSSM coupled to supergravity the gravitino obstructs it; that obstruction is not an inconsistency of the MSSM, only a statement that the symmetry cannot be gauged there. 
The number $16$ appears three times: as the topological superconductor's $\Z_{16}$, as the Standard Model's sixteen fermions, and now in the $8\times7=56$ ledger of $\Neq{8}$. These are linked by anomaly matching across the Smith homomorphism of Eq.~\eqref{eq:smith}. We stress that anomaly matching is a \emph{necessary} condition for a duality, not a sufficient one.
The topological superconductor and maximal supergravity live in different dimensions with unrelated dynamics, and the Smith homomorphism relates their anomaly theories, not the theories themselves.
It also resonates with the swampland philosophy that a consistent theory of quantum gravity should carry no uncanceled bordism invariants~\cite{McNamara:2019rup}.
Here the would-be $\Z_{16}$ obstruction is retired by the gravity multiplet itself, without added matter. Concretely, $\nu=0$ certifies that the $\mathcal{N}=8$ partition function is unambiguous on $\Spin^{\Z_4}$ spacetimes. Whether gauging this $\Z_4$ yields a new global form of maximal supergravity, and how its uniqueness at $\mathcal{N}=8$ feeds the cobordism program relating gravity to symmetry-protected phases, we leave as open questions.

Several caveats sharpen rather than soften the result. 
The relevant $\Z_4$ must be a genuine (gauged or at least well-defined) symmetry with $\Z_2=(-1)^F$; we have shown the anomaly vanishes for all such choices, but which is dynamically realized depends on the completion.
The gauged $SO(8)$ theory deserves a dedicated analysis. 
And Eq.~\eqref{eq:seven} inherits the $\RP^5$ conventions of Ref.~\cite{GarciaEtxebarria:2018ajm}. None of these undermines the central arithmetic of Eq.~\eqref{eq:main}.
Debray and Yu's~\cite{Debray:2022wcd} independent gravitino computation on the Wu manifold uses a formally different, dimension-dependent Rarita-Schwinger ghost-subtraction coefficient than Eq.~\eqref{eq:seven}.
Reconciling the two conventions would give $w_{3/2}=7$, a fully independent cross-check.
Finally, it would be interesting to know whether other extended supergravities, or string compactifications preserving them, share this feature, and whether the correspondence extends to the mixed and gravitational sectors beyond the $\Z_{16}$ studied here.

\begin{acknowledgments}
We thank Mikhail Solon for encouragement, and Prarit Agarwal, Arun Debray, Miguel Montero, Andreas Stergiou, and Matthew Yu for feedback on the Letter.
The computations reported here and drafts of this Letter were produced with the assistance of Claude Opus~4.8 (Anthropic). 
CWM's access to Claude Code was funded by Klaviyo.
Our \texttt{daifreed} Python package~\cite{Murphy:2026} depends on \texttt{SymPy}~\cite{10.7717/peerj-cs.103} and \texttt{mpmath}~\cite{mpmath}.
\end{acknowledgments}

%======================================================================
\bibliographystyle{apsrev4-2}
\bibliography{refs}

%======================================================================
\clearpage
\onecolumngrid
\newpage

\section{Supplemental Material: the \texttt{daifreed} library for Dai-Freed anomalies}
Every number quoted in the Letter is produced by \texttt{daifreed}, a small, tested open-source Python library that implements the Dai-Freed anomaly machinery of Ref.~\cite{GarciaEtxebarria:2018ajm}. 
Here we sketch what it does and give the short scripts that reproduce our results. 
However, the library is far more general than the $\Neq{8}$ application.
It also reproduces, e.g., the Standard Model and Grand Unified Theory anomaly-freedom, the baryon-triality mod-9 anomaly, and the Iba\~nez-Ross comparison.

Following Ref.~\cite{GarciaEtxebarria:2018ajm}, a Dai-Freed anomaly of a 4d theory with symmetry structure is a homomorphism $\Omega^{\text{struct}}_5(BG)\to U(1)$, computed in two steps: (i) find the bordism group; (ii) evaluate $\exp(2\pi i\,\eta)$ on generators. 
The library provides an exact $\eta$-invariant engine (the Bahri-Gilkey~\cite{bahri1987} lens-space/spherical-space-form formulas), the closed-form cancellation conditions derived from them, tabulated bordism groups, and the topological-superconductor ($\Z_4$, mod-16) module used here.

$\eta$-invariants of lens spaces are finite sums over roots of unity that, by bordism invariance, evaluate to exact rationals mod 1. 
The engine sums them symbolically in \texttt{SymPy} and recovers the rational by a \emph{certified} continued-fraction recognition (with an optional \texttt{mpmath} high-precision fast path; the two agree). 
No floating-point tolerances enter the reported results.

The mod-16 module weights each fermion by $\chi(q)$, Eq.~\eqref{eq:chi}, and by a spin-dependent factor: $1$ for a Weyl fermion, and $7$ for a gravitino. 
The factor $7$ is \emph{derived} (\code{rarita\_schwinger\_factor} returns the vector-spinor dimension $6$ plus the ghost $1$), reproducing Eq.~\eqref{eq:seven}. 
The library currently applies this weight uniformly across $\Z_4$ charge $q$; an independent Spin$^c$-embedding derivation of that charge-independence, of the kind used for the ordinary spin-$\tfrac12$ conditions in \code{daifreed.conditions}, is not yet implemented and is left for future work.

The central result, Eq.~\eqref{eq:main}:
\begin{lstlisting}
from daifreed import models
from daifreed.topological_sc import topological_superconductor

n8 = models.n8_supergravity()          # 8 gravitini (spin 3/2) + 56 spin-1/2
print(topological_superconductor(n8, "Z4"))
# TopSuperconductorResult(is_free=True, index_mod16=0, majorana_count=64, ...)
\end{lstlisting}
Here \code{majorana\_count} counts fermion \emph{fields} of odd $\Z_4$ charge ($8+56=64$)
It is not the domain-wall Majorana-\emph{mode} count of the discussion above, which weights each gravitino by $7$ and would give $8\times7+56=112$. 
Both are $\equiv0\pmod{16}$, but the two quantities differ and should not be conflated.

Anomaly-freedom for every $\Z_4\subset SU(8)$ with $\Z_2=(-1)^F$, Table~\ref{tab:enum}:
\begin{lstlisting}
from daifreed.topological_sc import z4_index
for p in (0, 2, 4, 6, 8):                # p = number of +i eigenvalues on the 8
    print(p, z4_index(models.n8_supergravity_z4(p), "Z4"))
# 0 0 / 2 0 / 4 0 / 6 0 / 8 0   -> all anomaly-free
\end{lstlisting}

The mixed $\Z_4$--gauge tool, validated on the MSSM (it reproduces the nonzero $SU(3)$-- and $SU(2)$--$\Z_4$ anomalies of Ref.~\cite{GarciaEtxebarria:2018ajm}):
\begin{lstlisting}
from daifreed.spectrum import Fermion, Spectrum
from daifreed.topological_sc import mixed_z4_gauge_anomaly

# SU(3)-Z4 from gauginos (adjoint index 2N = 6): anomalous.
su3 = Spectrum([Fermion({"Z4": 1, "idx": 6}, 1)])
print(mixed_z4_gauge_anomaly(su3, "Z4", "idx").is_free)   # False
\end{lstlisting}

The library ships with a test suite that pins each reproduced result, including all of the above and the Bahri-Gilkey identity $\eta_{s,q}(L^1(n))=-s/n \bmod 1$. 
The repository is publicly available at \url{https://gitlab.com/chris-w-murphy/dai-freed}, and it includes instructions for using the code with other models.
Additionally, it can be installed from the Python Package Index e.g.\ by running \texttt{uv add daifreed} or \texttt{pip install daifreed} in a command-line interface.

%======================================================================
\end{document}